\documentclass[aps,twocolumn,pra,floatfix]{revtex4-2}
\usepackage{epsfig}
\usepackage{graphicx,subfigure}

\usepackage{dcolumn}
\usepackage{amssymb,amsmath}
\usepackage{mathrsfs}

\usepackage{soul}

\usepackage{bm,amsfonts,mathtools}

\usepackage{tikz}
\usepackage{xcolor}

\begin{document}

	\title{Prospects for optical clocks combining high sensitivity to new physics with insensitivity to perturbations; the case of Sb$^{+}$, Au$^{+}$, and Hg$^{2+}$
	}

	\author{Saleh O. Allehabi$^{1,2}$}
	
	\author{V. A. Dzuba$^1$}
	\author{V. V. Flambaum$^{1}$}

	\affiliation{$^1$School of Physics, University of New South Wales, Sydney 2052, Australia}
	
	\affiliation{$^{2}$Department of Physics, Faculty of Science, Islamic University of Madinah, Madina 42351, Kingdom of Saudi Arabia}

	\date{\today}
	
	\begin{abstract}
		
Our study is motivated by the prospect of several metastable states in the Sb$ ^{+} $, Au$ ^{+} $, and Hg$ ^{2+} $ ions being used as possible candidates for optical clocks. We calculate several atomic properties relevant to the development of optical clocks for those clock transitions using two different approaches of relativistic many-body calculations, configuration interaction with single-double
coupled-cluster (CI+SD) method, and configuration interaction with perturbation theory (CIPT) method. Our results demonstrate that the relative black body radiation shifts for these transitions are small, $ \sim 10 ^{-16} $. It is also found that there is considerable sensitivity to new physics, as evidenced by a significant enhancement of the effect of the time variation of the fine structure constant $\alpha$ on the frequencies of the clock transitions. The corresponding factor ranges from -5.56 to 2.20.
Our results are compared with available previous data.
	\end{abstract}
	
	\date{\today}
	
	\maketitle

	\section{Introduction}

Presently the fractional precision of the frequency measurements in optical clock transitions of Sr, Yb, Al$ ^{+} $, Hg, Hg$ ^{+} $, and Yb$ ^{+} $ atomic systems reached an unprecedented level of 10$ ^{-18} $~\cite{brewer2019al+,ludlow2018optical,huntemann2016single,mcgrew2018atomic,masao2015cryogenic}. 
This greatly  benefits the search for new physics (NP) beyond the standard model. One possible manifestation of NP is  the change of the transition frequencies in time due to, e.g., time variation of the fine structure constant $\alpha$. 
The majority of operational atomic optical clocks, however, are not sensitive to this variation. If we use the so-called {\em enhancement factor} $K$ defined as a link between  the time variation of atomic frequency and time variation of $\alpha$, $\dot \omega/\omega = K \dot \alpha/\alpha$, then for a majority of operational optical clock transitions $K<1$. For example, $K$ = 0.008 for Al~II, $K$ = 0.15 for Ca~II, $K$ = 0.06 for Sr~I, and $K$ = 0.31 for Yb~I~\cite{dzuba2009atomic}.
The exceptions are clock transitions in Yb~II and Hg~II which have enhancement factors $K=-5.3$ and $K=-2.94$ respectively~\cite{dzuba2009atomic}.
These transitions have already been used in the search for time variation of $\alpha$ producing the limit on the yearly fractional change on the level $\sim 10^{-18}$~\cite{filzinger2023improved}.
	
A number of promising optical clock transitions sensitive to variation of $\alpha$ were suggested in our earlier works~\cite{dzuba2018testing,dzuba2021time,allehabi2021using,allehabi2022atomic}.
In the present work, we continue the search by studying metastable states of Sb$^{+}$, Au$^{+}$, and Hg$^{2+}$ ions.
Each of these systems has several metastable states which bring extra possibilities. One could combine a pair of frequencies to suppress black body radiation (BBR) shift~\cite{Yudin}. Also, measuring one combination of frequencies against the other allows for monitoring time variation of the fine structure constant. In addition, mercury has seven stable isotopes, five of which  have zero nuclear spin. This makes the Hg$^{2+}$ ion a good candidate for the study of King plot nonlinearities. Having four stable isotopes and two transitions with a high level of accuracy of the measurement are the minimum requirements for such a study. It may bring information concerning the nuclear structure~\cite{berengut2018probing,flambaum2018isotope}
and helps to put constrain on the strength of new interactions mediated by scalar bosons~\cite{boson1,boson2,boson3}.
		
To perform the calculations, we use the configuration interaction with single-double coupled clusters (CI + SD)~\cite{dzuba2014combination} method for the Sb$^{+}$ ion and the configuration interaction with perturbation theory (CIPT)~\cite{dzuba2017combining} method for the Au$^{+} $, and Hg$^{2+} $ ions.
Energy levels, Land\'{e} $g$-factors, transition amplitudes (E1, M1, E2, and hyperfine induced) between low-lying states, quadrupole moments, and lifetimes of clock states have been calculated. 
We also calculated the scalar polarizabilities of the ground and excited clock states in order to estimate the blackbody radiation shifts of the clock frequencies. Calculating clock frequencies with the different values of the fine structure constant $\alpha$ in computer codes reveals the sensitivity of the transitions to the $\alpha$-variation.

\section{Method of Calculations}
	
\subsection{Calculation of energy levels}

The choice of computational method depends on the number of external (valence) electrons, i.e. electrons outside of the compact closed-shell core.	
Sb$^{+} $ has four valence electrons, its ground state configuration is [Pd]$5s^25p^2$.
Both, Au$ ^{+} $ and Hg$ ^{2+} $ ions have closed-shell ground states of the [Yb]$5d^{10}$ configuration.
However, all excited states of these ions have excitations from the $5d$ subshell. Therefore, it is logical to consider these systems as having open $5d$ subshell, i.e. having ten external electrons. Two different methods are used to treat these very different systems. We use the CI+SD (configuration interaction with single-double coupled-cluster) method~\cite{dzuba2014combination} for Sb$^{+} $. It includes the accurate CI treatment of the correlations between four valence electrons and the SD treatment of the core-core and core-valence correlations. 
Such an approach gives pretty accurate results for systems with a relatively simple configuration of external electrons (up to four electrons). 
It cannot be used  for the Au$^{+} $ and Hg$^{2+} $ ions because of the large number of external electrons. Here we use the CIPT (configuration interaction with perturbation theory)  method~\cite{dzuba2017combining} which was especially developed for such systems. It is less accurate. It does not include correlations between external electrons and core electrons below the $5d$ subshell, and part of the correlations between external electrons are included perturbatively . However, it gives reasonably good results for low states of atomic systems with complicated electronic structures.

\subsubsection{The CI+SD method}
	
As a starting point, we use the relativistic Hartree-Fock (RHF) calculations for a closed-shell core based on the $V^{N-M} $ approximation (where $ N $ is the total number of electrons, and $ M $ is the number of valence electrons)~\cite{dzuba2005v}. The RHF Hamiltonian can be written as	
\begin{equation} 
	\hat H^{\rm RHF}= c\bm{\alpha}\cdot\mathbf{p}+(\beta -1)mc^2+V_{\rm nuc}(r)+V_{\rm core}(r),
\end{equation}
where $c$ is the speed of light, $\bm{\alpha}$ and $\beta$ are Dirac matrices, $\bm{p}$ represents the electron momentum, $m$ is the mass of electron, $V_{\rm nuc}$ is the nuclear potential derived by integrating the Fermi distribution of the nuclear charge density, and $V_{\rm core}(r)$ is the self-consistent RHF potential which is formed by the electrons of the closed-shell core.
	
After completing the self-consistent procedure for the core, a complete set of single-electron basis wave functions is calculated using the B-spline technique~\cite{johnson1986computation,johnson1988finite}. Each basis state is a linear combination of B-splines which is an eigenstate of the RHF Hamiltonian.
The basis set is built up of 40 B-splines of order 9 in a box that has a radius $R_{\rm max}=40a_B$, where $a_B$ is the Bohr radius, with the orbital angular momentum 0~$\leq$~\textit{l}~$\leq$~6. With these parameters ($l_{\rm max}$, $ R_{\rm max}$, and the number of B-splines), 
the saturation of the basis for low-lying many-electron states  is achieved. 
Increasing the values of the parameters leads to insignificant changes.
	
The SD equations are first solved for the core, leading to the inclusion of the core-core correlations. Then the SD equations are solved for valence states.
As a result, the correlation operators $\Sigma_1$ and $\Sigma_2$ are formed~\cite{dzuba1987correlation,dzuba1996combination,dzuba2005v}. 
$\Sigma_1$ describes the correlation interaction between a particular valence electron and the core, whereas $\Sigma_2$ describes the Coulomb interaction screening between a pair of valence electrons. 


	
The effective CI+SD Hamiltonian has the form
\begin{eqnarray} \label{e:HCI}
		\hat H^{\rm eff}=\sum_{i=1}^{M} \left(\hat H^{\rm RHF}+\Sigma_1\right)_i 
		+\sum_{i<j}^{M} \left(\dfrac{e^2}{|r_i-r_j|}+ \Sigma_{2ij}\right)
\end{eqnarray}
Here $i$ and $j$ enumerate valence electrons, and summation goes over valence electrons.

\begin{figure*}[tb]
	\centering
	\subfigure[~~Sb$ ^{+} $]{\includegraphics[width=0.32\textwidth]{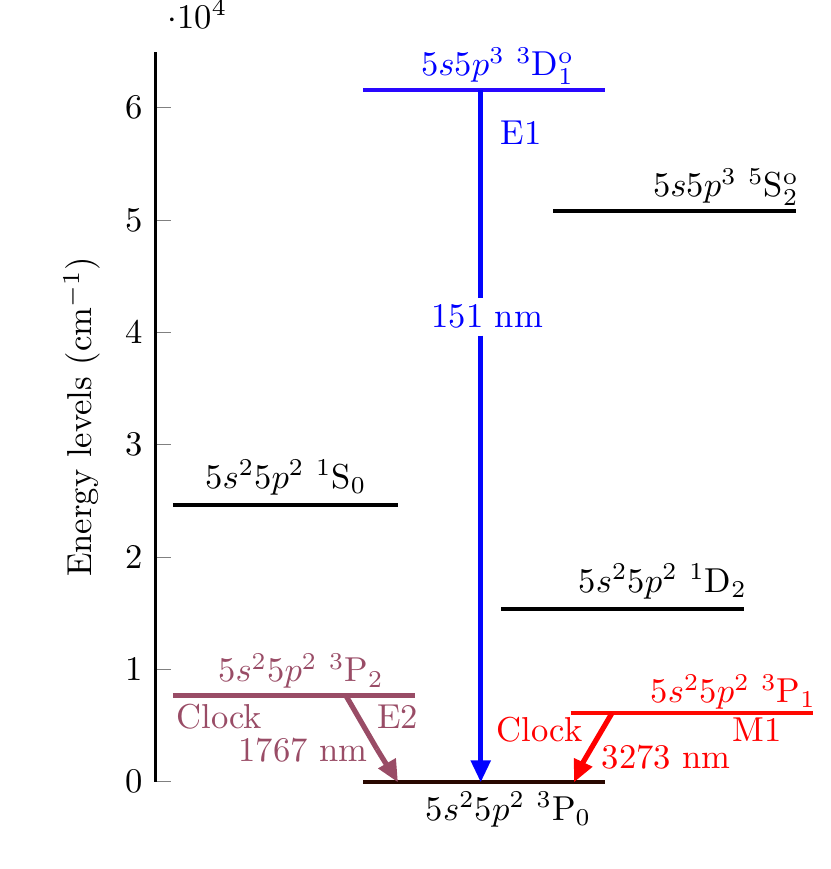}} 
	\subfigure[~~Au$ ^{+} $]{\includegraphics[width=0.32\textwidth]{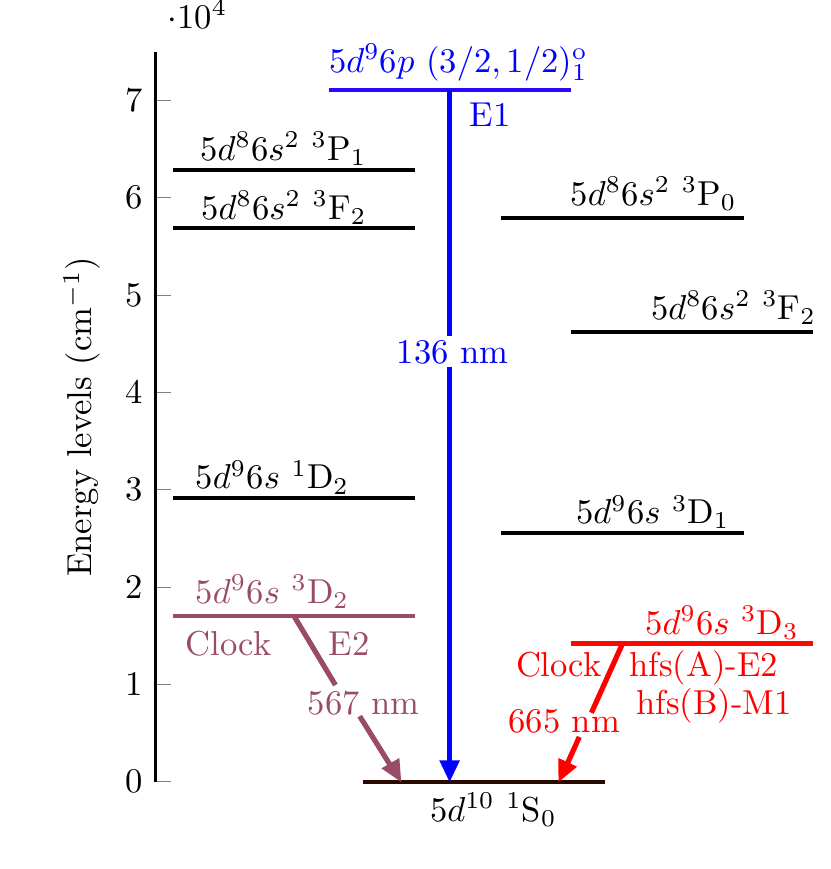}} 
	\subfigure[~~Hg$ ^{2+} $]{\includegraphics[width=0.32\textwidth]{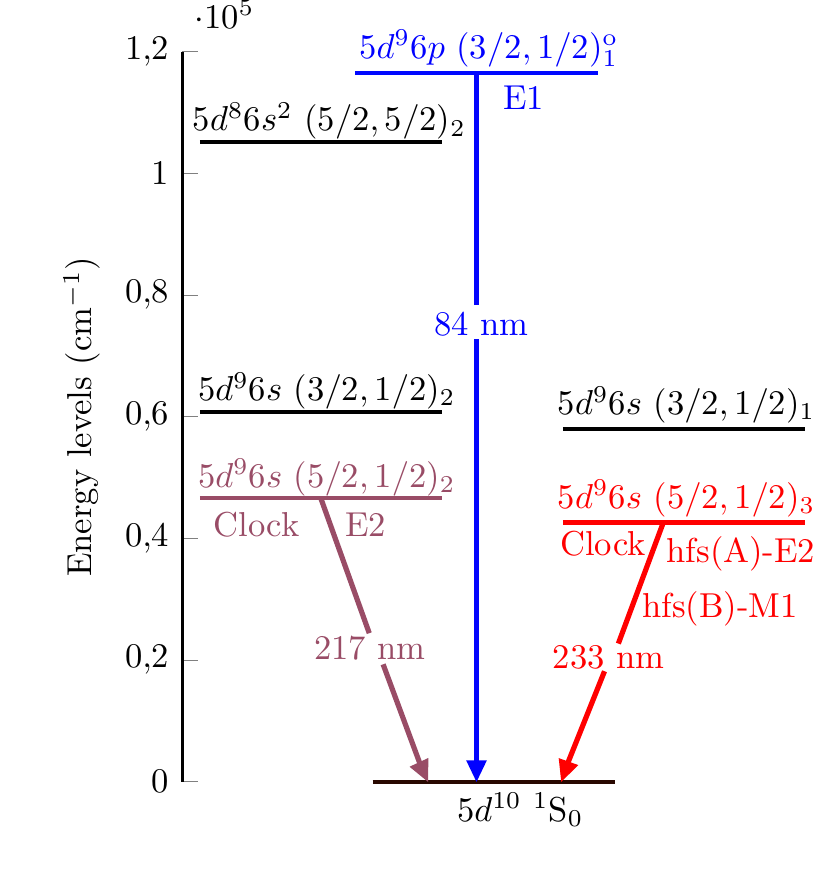}} 
	
\caption{The energy diagram for the states of the Sb$^{+} $, Au$^+ $, and Hg$^{2+} $ ions is relevant for the optical ion clocks. The clock transitions are shown as red, brown, and green lines. The blue color depicts the first odd state connected to the ground state via E1 transition. Due to the fact that they have many decay channels to lower-lying excited states, they cannot be used for cooling purposes.}

\label{f:1}
\end{figure*}

	\subsubsection{ The CIPT method}
	
	
As previously mentioned, this method has been developed particularly for atomic systems with unfilled shells that have a large number of electrons above the closed-shell core. The vital characteristic of this method is that the off-diagonal matrix elements between high excited basis states are neglected in order to reduce the CI matrix to an effective matrix of small size with perturbative contributions from high states.
Thus, for the price of some sacrifices in accuracy, the problem becomes manageable.
	
The calculations start from the RHF calculation for the closed-shell ground states of the Au$^{+} $ and Hg$^{2+} $ ions.
The $ V^{N }$ potential and B-spline technique are used to calculate a complete set of single-electron basis states.
We use the same set of parameters as in the previous method, with the only difference being the maximum angular moment ($l_{\rm max}$), for which we use $l_{\rm max}=4$. Note that $l_{\rm max}=6$ is needed for accurate calculation of the $\Sigma_1$ and $\Sigma_2$ operators. In the CIPT calculations, we don't have these operators and $l_{\rm max}=4$ is sufficient for the CI calculations.
	
The wave function for valence electrons	$\Psi$ is written as an expansion over single-determinant many-electron basis functions $\Phi_{i}$.
The expansion is divided into two parts
\begin{equation} \label{e:Wave}
	\begin{aligned}
		\Psi\left(r_{1}, \ldots, r_{M}\right)=& \sum_{i=1}^{N_{\text {eff }}} c_{i} \Phi_{i}\left(r_{1}, \ldots, r_{M}\right) \\
		&+\sum_{i=N_{\text {eff }}+1}^{N_{\text {total}}} c_{i} \Phi_{i}\left(r_{1}, \ldots, r_{M}\right) 
	\end{aligned}
\end{equation}
where $c_{i}$ are the expansion coefficients. A relatively small number of the low-energy terms constitutes the first part of the expansion and is considered to be a good approximation to the wave function.
A large number of high-energy states are included in the second part, which is considered as a minor correction to the valence wave function ($N_{\text {eff }}$ $<$ $i$ $\leqslant$ $N_{\text {total}}$, where $N_{\text {total }}$is the total number of the basis states). 
Neglecting the off-diagonal matrix elements between terms in the second summation in Eq.~(\ref{e:Wave}) $\left(\left\langle i\left|H^{\mathrm{eff}}\right| h\right\rangle=0 \text { for } N_{\text {eff }}< i, h \leqslant N_{\text {total}}\right)$ allows to reduce the total CI matrix to the effective CI matrix of much smaller size
\begin{equation}
		\langle i|H^\mathrm{eff}|g\rangle\rightarrow\langle i|H^\mathrm{eff}|g\rangle+\sum_{k}\frac{\langle i|H^\mathrm{eff}|k\rangle\langle k|H^\mathrm{eff}|g\rangle}{E-E_{k}}.
		\label{e:CIPT}
\end{equation}
Here, $i, g \leqslant N_{\text {eff }}, N_{\text {eff }}<k \leqslant N_{\text {total}}$, $E$ is the energy of the state of interest, and $E_k$ denotes the diagonal matrix element for high-energy states, $E_{k}=\left\langle k\left|H^{\mathrm{eff}}\right| k\right\rangle$. The summation in (\ref{e:CIPT}) runs over all high-energy states. 
Note that the parameter $E$ in the denominator of (\ref{e:CIPT}) is the same as the energy of the state of interest which is to be obtained from solving the CI equations. Since this energy is not known in advance, iterations over $E$ are needed to find it.
 A more comprehensive explanation of the method can be found in Ref.~\cite{dzuba2017combining}.

\subsection{Calculation of transition amplitudes and lifetimes}

Transition amplitudes are calculated 	using the time-dependent Hartree-Fock (TDHF) method, which is equivalent to the well-known random phase approximation (RPA).
The RPA equations for the core are
\begin{equation}\label{e:RPA}
	\left(\hat H^{\rm RHF}-\epsilon_c\right)\delta\psi_c=-\left(\hat f+\delta V^{f}_{\rm core}\right)\psi_c,
\end{equation}
Index $c$ numerates states in the core, $\epsilon_c$ is the energy of electron state $ c $, and $\psi_c$ is the state's wave function, $\delta\psi_c$ is a correction to the wave function due to the external field $\hat f$, 
$\delta V^{f}_{\rm core}$ is the correction to the self-consistent RHF potential of the core 
generated by the modifications to all core states in the external field. The RPA equations (\ref{e:RPA}) are solved self-consistently for the sake of finding $\delta V^{f}_{\rm core}$.
After that, the transition amplitude between valence states $a$ and $b$ is given by the off-diagonal matrix element
\begin{equation}\label{e:Ab}
	A_{a b}=\left\langle b\left|\hat f+\delta V^{f}_{\rm core }\right| a\right\rangle.
\end{equation}
Here, $|a\rangle$ and $|b\rangle$ are the many-electron wave functions calculated with the methods described above.
		

The probabilities of spontaneous emission due to E1, M1, and E2 transitions from upper state $b$ to lower state $a$ are given by (in atomic units).
	
\begin{equation}\label{e:Td}
	T_{E1,M1} = \frac{4}{3}(\alpha\omega_{ab})^3 \frac{A^2_{E1,M1}}{2J_b+1},
\end{equation}
	
\begin{equation}\label{e:Tq2}
	T_{\rm E2} = \frac{1}{15}(\alpha\omega_{ab})^5 \frac{A^2_{E2}}{2J_b+1},
\end{equation}
Here $\alpha$ is the fine structure constant ($\alpha\approx\frac{1}{137}$), $\omega_{ab}$ is the energy difference between the lower and upper states, \textit{A} is the transition amplitude (reduced matrix element) (\ref{e:Ab}), and $J_b$ is the total angular momentum of the upper state \textit{b}. Note that magnetic amplitudes $A_{M1}$ contain the Bohr magneton $\mu_B$ ($\mu_B = \alpha/2 \approx 3.65 \times 10^{-3}$ in the Gauss-type atomic units). 
	
The lowest metastable states of Au$ ^{+} $ and Hg$ ^{2+} $ have the same parity as the ground states but the value of the total angular momentum $J$ is different by 3. This means that dominating transition channel in Au$ ^{+} $ and odd isotopes of Hg$ ^{2+} $ comes from E2 or M1 electromagnetic transition induced by hyperfine structure (hfs). More specifically, it is a magnetic dipole (M1) transition in combination with the electric dipole hfs for $^{197}$Au$^+$ and $^{201}$Hg$^{2+}$ or electric quadrupole (E2) transition in combination with magnetic dipole hfs for $^{197}$Au$^+$, $^{199}$Hg$^{2+}$ and $^{201}$Hg$^{2+}$. The transition amplitude is given by
\begin{eqnarray}\label{e:hfi}
			A_{\mathrm{(H-T)}}(a \leftrightarrow b)&= & \sum_{n}\left(\frac{\left\langle a\left|A_{\mathrm{H}}\right| {n}\right\rangle\left\langle {n}\left|A_{\mathrm{T}}\right| b\right\rangle}{E_{a}-E_{n}}\right) \\
			& +&\sum_{n}\left(\frac{\left\langle a\left|A_{\mathrm{T}}\right| {n}\right\rangle\left\langle {n}\left|A_{\mathrm{H}}\right| b\right\rangle}{E_{b}-E_{n}}\right), \nonumber
\end{eqnarray}
where $ A_{\mathrm{H}} $ is the operator of the magnetic dipole (A) or electric
quadrupole (B) hfs interaction, and $ A_{\mathrm{T}} $ is the  operator  of the
E2 or M1 transitions. 
Summation goes over the complete set of intermediate states.
In our considered transition ($ ^{1}$S$_{0} - ^3$D$_{3} $), we include three states ($ ^{3} $D$ _{2} $, $ ^{3} $D$ _{1} $, and $ ^{1} $D$ _{2} $) which give dominating contributions.

Accordingly, for the hfs(A)-E2 mechanism, the equation above (Eq. (\ref{e:hfi})) would read as follows
\begin{equation} 
	\begin{aligned}\label{e:AE2}
		&	A_{\mathrm{H_{A}}-\mathrm{E} 2}(^3{\rm D}_3 \leftrightarrow ^1{\rm S}_0)=\\&
		\frac{\left\langle ^3{\rm D}_3\left|A_{\mathrm{H_{A}}}\right| {^3{\rm D}_2}\right\rangle\times\langle  ^3{\rm D}_2|A_{E2}| ^1{\rm S}_0\rangle}{E_{^3{\rm D}_3}-E_{^3{\rm D}_2}}+
		\\& 
		\frac{\left\langle ^3{\rm D}_3\left|A_{\mathrm{H_{A}}}\right| {^1{\rm D}_2}\right\rangle\times\langle ^1{\rm D}_2|A_{E2}| ^1{\rm S}_0\rangle}{E_{^3{\rm D}_3}-E_{^1{\rm D}_2}}+
		\\&
		\frac{\left\langle {^3{\rm D}_3}|A_{E2}| ^3{\rm D}_1  \right\rangle\times\langle ^3{\rm D}_1\left|A_{\mathrm{H_{A}}}\right|{^1{\rm S}_0}\rangle}{E_{^3{\rm D}_1}-E_{^1{\rm S}_0}}
	\end{aligned}
\end{equation}
Likewise, for the hfs(B)-M1 mechanism, the equation is
\begin{equation} 
	\begin{aligned}\label{e:BM1}
		&	A_{\mathrm{H_B}-\mathrm{M} 1}(^3{\rm D}_3 \leftrightarrow ^1{\rm S}_0)=\\&
		\frac{\left\langle ^3{\rm D}_3\left|A_{M1}\right| {^3{\rm D}_2}\right\rangle\times\langle ^3{\rm D}_2|A_{\mathrm{H_B}}| ^1{\rm S}_0\rangle}{E_{^3{\rm D}_2}-E_{^1{\rm S}_0}}+
		\\& 
		\frac{\left\langle ^3{\rm D}_3\left|A_{M1}\right| {^1{\rm D}_2}\right\rangle\times\langle ^1{\rm D}_2|A_{\mathrm{H_B}}| ^1{\rm S}_0\rangle}{E_{^1{\rm D}_2}-E_{^1{\rm S}_0}}+
		\\&
		\frac{\left\langle ^3{\rm D}_3|A_{\mathrm{H_B}}| {^3{\rm D}_1} \right\rangle\times\langle {^3{\rm D}_1}\left|A_{M1}\right| ^1{\rm S}_0 \rangle}{E_{^3{\rm D}_3}-E_{^3{\rm D}_1}}
	\end{aligned}
\end{equation}
Then, transition rates can be calculated using Eq. (\ref{e:Tq2}) for hfs(A)-E2 and Eq. (\ref{e:Td}) for hfs(B)-M1.

In order to determine the lifetime ($\tau_b$, expressed in seconds) of each excited state $b$, the following formula is used:
\begin{equation}\label{e:tau}
	\tau_b =  2.4189 \times 10^{-17}/\sum_a T_{ab},
\end{equation}
where the summation goes over all the posible transitions to lower states $a$.

\begin{table*}
		
	\caption{\label{t:Energy}
			Excitation energies ($E$), Land\'{e} \textit{g}-factor, and lifetimes ($\tau$) for the first excited states.} 
		\begin{ruledtabular}
			\begin{tabular}{cc cc cc cc cr}
				&&&&
				
				\multicolumn{2}{c}{E [cm$^{-1}$]}&
				\multicolumn{2}{c}{\textit{g}-factor}&
				
				\multicolumn{2}{c}{$\tau$ [s]}\\
				\cline{5-6}
				\cline{7-8}
				\cline{9-10}
				\multicolumn{1}{c}{No.}& 
				\multicolumn{1}{c}{Conf.}&
				\multicolumn{1}{c}{Term}&
				
				\multicolumn{1}{c}{$J$}&
				\multicolumn{1}{c}{Present}&
				\multicolumn{1}{l}{NIST~\cite{kramida2019nist}}&
				\multicolumn{1}{c}{Present}&
				\multicolumn{1}{l}{NIST~\cite{kramida2019nist}}&
				\multicolumn{1}{c}{Present}&
				\multicolumn{1}{c}{Other}\\
				\hline
				
				\multicolumn{5}{c}{\textbf{Sb~II}}&
				\multicolumn{1}{c}{}&&&&\\

				1 & $5s^25p^2$& $^3${P}& {0} &0 &0&0.000&&&\\
				2 & $5s^25p^2$& $^3${P}& {1} &3146&3054.6&1.500&&1.38 s& \\
				3& $5s^25p^2$& $^3${P}& {2} &5504 &5658.2&1.406&&5.09 s& \\
				4& $5s^25p^2$& $^1${D}& {2}&12966 &12789.8&1.094&&&\\
				5& $5s^25p^2$& $^1${S}& {0}&24437&23905.5&0.000&&&\\

				\hline
				
				\multicolumn{5}{c}{\textbf{Au II}}&
				\multicolumn{1}{c}{}&&&&\\

				1& $5d^{10}$& $^1${S}& {0}&0&0&0.000&&&\\
				
				2& $5d^{9}6s$& $^3${D}& {3}&15095&15039.572&1.333&&$25 \times 10^8$  s\tablenotemark[1]&\\
				
				3& $5d^{9}6s$& $^3${D}& {2}&17832  &17640.616&1.105&&1.33 s&\\
				
				4& $5d^{9}6s$& $^3${D}& {1}&27225&27765.758 &0.500&&37.9 ms&\\
				
				5& $5d^{9}6s$& $^1${D}& {2}&29817& 29621.249 & 1.062&  1.04 	&&\\
			\hline
				
			\multicolumn{5}{c}{\textbf{Hg III}}&
			\multicolumn{1}{c}{}&&&&\\
				
			 1& 5$ d^{10}  $  & $^{1}$S         & 0 & 0.00     & 0.00&0.000&& &       \\ 
				2& $ 5d^{9}6s $  & (5/2,1/2)  & 3 & 43360 &42850.3&1.333&& $9 \times 10^5$ s\tablenotemark[2]& \\ 
				3& $ 5d^{9}6s $  & (5/2,1/2)  & 2 & 46616&46029.5&1.099&&28.7 ms &(34$\pm$3 ms)$_{\rm expt.}$~\cite{calamai1990radiative}   \\ 
			4& $ 5d^{9}6s $  & (3/2,1/2)  & 1 &  58389& 58405.8&0.500&&& \\ 
			5& $ 5d^{9}6s $  & (3/2,1/2)  & 2 &  61379&  61085.7&1.064&&& \\ 				
			\end{tabular}
		\tablenotetext[1]{The lifetime of $^{197}$Au$ ^{+} $ isotope.} 
			\tablenotetext[2]{The lifetime of $^{201}$Hg$ ^{2+} $ isotope.}
		\end{ruledtabular}
		
\end{table*}

\begin{table*}
		
\caption{\label{t:Tran} Transition amplitudes ($A$, a.u.) and transition probabilities ($T$, 1/s) evaluated with NIST frequencies for some low  states.
HFS-induced transitions are calculated for $^{197}$Au ($I=3/2$, $\mu=0.1451$~\cite{kramida2019nist}, $Q=0.547(16)$~\cite{Stone}) and $^{201}$Hg ($I=3/2$, $\mu=-0.5602$~\cite{kramida2019nist}, $Q=0.387(6)$~\cite{Stone}). 
		Numbers in square brackets stand for powers of ten.} 
		
		\begin{ruledtabular}
			\begin{tabular}{lc ll ll l}
				&&
				\multicolumn{2}{c}{($\omega$), NIST~\cite{kramida2019nist} }&
				\multicolumn{2}{c}{Present}&
				\multicolumn{1}{c}{Ref.~\cite{biemont1995forbidden}}\\

				\cline{3-4}
				\cline{5-6}

				\multicolumn{1}{c}{Transition}& 
				\multicolumn{1}{c}{Type}&
				\multicolumn{1}{c}{ [cm$^{-1}$]}&
				\multicolumn{1}{c}{ [a.u.]}&

				\multicolumn{1}{c}{$A$ [a.u]}&
				\multicolumn{1}{c}{$T$ [s$^{-1}$]}&
				\multicolumn{1}{c}{$T$ [s$^{-1}$]}\\
				
				\hline
				
				\multicolumn{7}{c}{\textbf{Sb~II}}\\
				2 $-$ 1       & M1   & 3054.6&	0.0139	&5.02[-3]&	0.727 &0.500   \\

				3 $-$ 1      & E2   & 5658.2&	0.0258&	5.12&	3.40[-3] &3.73[-3]\\
				3 $-$ 2      & M1   &2603.6	&0.0119	&5.20[-3]&	0.193&0.192  \\
				3 $-$ 2      & E2   & 2603.6&	0.0119&	-6.51&	1.13[-4]&   1.18[-4] \\
				\hline

				\multicolumn{7}{c}{\textbf{Au II}}\\
								
				2 $-$ 1      & H$_{\rm A}$-E2 &15039.572&0.0685& 1.797[-4]&		 3.97[-10] & \\
				
					2 $-$ 1      & H$_{\rm B}$-M1  &15039.572&0.0685& 1.707[-9]	&  2.86[-12]& \\

				3 $-$ 1      & E2 &17640.616&	0.0804&	3.5113&	0.4713	& \\
				3 $-$ 2      & M1 &2601.044&	0.0119&	-0.00626&	0.279& \\
				3 $-$ 2      & E2   &2601.044&	0.0119&	1.77&	8.38[-6]	&  \\

				4 $-$ 1      & M1   &27765.758&0.1265&0.231[-5]&7.69[-5]&   \\

				4 $-$ 2      & E2   &12726.186&0.0580&0.683&5.80[-3]&  \\
				
				4 $-$ 3     & M1   & 10125.142&0.0461&-0.00614&26.4&  \\
				4 $-$ 3     & E2   & 10125.142&0.0461&1.5844&9.96[-3]&  \\

				\hline

\multicolumn{7}{c}{\textbf{Hg III}}\\
				
	2 $-$ 1      & H$_{\rm A}$-E2 &42850.3&0.1952& -6.872[-4]&  1.09[-6]& \\
				
	2 $-$ 1      & H$_{\rm B}$-M1&42850.3	 &0.1952& 6.409[-10]&  9.33[-12]& \\
	3 $-$ 1      & E2 &46029.5&	0.2097& $ - $2.746&   34.86&
				\\
	3 $-$ 2      & M1 & 3179.2&	0.0145&   $ - $6.15[-3]&	0.492&\\
	3 $-$ 2      & E2   & 3179.2&	0.0145&   1.41&   1.45[-5]&\\

\end{tabular}
\end{ruledtabular}
\end{table*}

\section{RESULTS AND DISCUSSION}
	
\subsection{Energy levels, g-factors, transition amplitudes, and lifetimes of the systems}
	
	In Fig. \ref{f:1}, we present some low-lying states of the considered ions. The first two excited states of each ion, which are metastable states, are considered as clock states. The lowest odd state, which is linked to the ground state by an electric dipole transition (E1), has a large number of decay channels to lower-lying excited states, making them unsuitable for cooling purposes. Alternatively, sympathetic cooling may be used~\cite{QLS}.
This is done by co-trapping the ions with other ions (called logic ions), which can be cooled by lasers.
For efficient sympathetic cooling one needs the charge-to-mass ratio of the clock ion to be close 
to that of the logic ion~\cite{SC12}. The most suitable logic ion for Sb$ ^{+} $ and Hg$ ^{2+} $ ions is Sr$ ^{+} $, while for Au$ ^{+} $ it is Hg$ ^{+} $.
In all three cases, the  charge-to-mass ratios are very close.


In Table~\ref{t:Energy}, we present our results for the energies, $g$-factors, and lifetimes for the clock and other states. To illustrate the accuracy of our results, we compare them with measurements, where available.
In terms of energy levels, excellent agreement is found between experiments and our theoretical values for all considered ions. 
The discrepancy does not exceed 5\%.
	
The results for transition amplitudes ($A$) and transition rates ($T$) are presented in Table~\ref{t:Tran}, along with experimental results whenever possible and earlier calculations. We calculate transition rates for all possible transitions between low-lying states.
Experimental frequencies from the NIST database were used for calculations of the transition rates. Our results are in agreement with those from previous data. Calculated values of $T$ are used in Eq.~(\ref{e:tau}) to derive the lifetimes of clock states for all ions.
The results are summarized in Table~\ref{t:Energy}, together with experimental and earlier calculations. 
There is good agreement between the results of the study and those found in previous studies.	
	
\subsection{Polarizabilities and blackbody radiation shifts}

\label{s:BBR}
	
To study the performance of prospective optical clocks, it is essential to consider the dynamic and static scalar polarizabilities of the atoms. 
Scalar polarizability $\alpha_v(0)$ determines the value of the blackbody radiation (BBR) shift that affects the clock state frequency. This shift represents the primary source of uncertainty. 
	
$\alpha_v(0)$ of an atomic system in state $v$ can be expressed as a sum over a complete set of states $n$ 
connected to state $v$ by the electric dipole ($E1$) transitions (we use atomic units)
\begin{equation}\label{e:pol}
		\alpha_v(0)=\dfrac{2}{3(2J_v+1)}\sum_{n}\frac{A_{vn}^2}{\omega_{vn}},
\end{equation}
where $A_{vn}$ are reduced matrix elements, $J_v$ is the total angular momentum of state $v$, and $\omega_{vn}$ is the frequency of the transition. Notations $v$ and $n$ refer to many-electron atomic states.
	
A practical way to express the polarizability of atomic systems is to combine the polarizability of the closed-shell core with the contribution of the valence electrons. The polarizability of the closed-shell core is given in the RPA approximation by
\begin{equation}\label{e:pcore}
	\alpha_v(0)=\frac{2}{3} \sum_c\left\langle v|\hat{d}| \delta \psi_c\right\rangle,
\end{equation}
where $\hat{d}$ is the operator of the electric dipole moment and $\delta \psi_c$ is the RPA correction to the core state $c$ [see Eq.~(\ref{e:RPA})]. The summation goes over all states in the core. 

The Au$ ^{+} $ and Hg$ ^{2+} $ ions are closed shell systems in the ground state, with the $5d^{10}$ configuration of the outermost subshell.
The polarizability of these ions is given by (\ref{e:pcore}).
The Sb$ ^{+} $ ion is an open-shell system with the similar closed-shell core ($4d^{10}$) and four more electrons above closed shells.

\begin{table}
\caption{\label{t:CorPol}
Scalar static polarizabilities for the closed-shell core of the considered ions ($\alpha_0$) in atomic unit (a.u.).}
\begin{ruledtabular}
\begin{tabular}{ c cc c r}
&
\multicolumn{1}{c}{State}&
\multicolumn{3}{c}{$\alpha_{0}({\rm Core})$}\\
				
\cline{3-5}
				
				
\multicolumn{1}{c}{Ion} &
\multicolumn{1}{c}{closed-shell core} &
\multicolumn{1}{c}{Present} &
\multicolumn{1}{c}{Prev. Work} &
\multicolumn{1}{c}{Refs.}\\
				
\hline

Sb II&$4d^{10}$ &1.680&1.68
&\cite{hati1995electronegativity}$ _{\rm Expt.} $\\
				
&&&1.68& \cite{johnson1983electric}$ _{\rm Theo.} $\\
				
Au II&$5d^{10}$&12.40&12.705&\cite{hati1995electronegativity}$ _{\rm Expt.} $\\
				
Hg III&$5d^{10}$&7.376&8.401 &\cite{hati1995electronegativity}$ _{\rm Expt.} $\\
				
&&&6.421,7.056&\cite{kello1995polarized}$ _{\rm Theo.} $\\
				
\end{tabular}
\end{ruledtabular}
\end{table}

\begin{table*}
\caption{\label{t:pol}
Scalar static polarizabilities of the ground states, $\alpha_0({\rm GS})$, and clock states, $\alpha_0({\rm CS})$,  and BBR frequency shifts for the clock transitions.  $\delta\nu_{BBR}$/$\omega$ is the fractional contribution of the BBR shift; where $\omega$ is the clock transition frequency. The values presented for $\alpha_0({\rm GS})$ and $\alpha_0({\rm CS})$ are summaries of the total scalar polarizability (core + valence).
		}
		\begin{ruledtabular}
			\begin{tabular}{cc c cr ll}
				&&&

				\multicolumn{1}{c}{$\Delta \alpha (0)$} &
				\multicolumn{3}{c}{BBR, (\textit{T}= 300 K)} \\

				\cline{4-4}
				\cline{5-7}
				
				\multicolumn{1}{c}{Transition}&
				
				\multicolumn{1}{c}{$\alpha_0({\rm GS})$[$a_B^3$]}&
				\multicolumn{1}{c}{$\alpha_0({\rm CS})$[$a_B^3$]}&

				\multicolumn{1}{c}{$\alpha_0({\rm CS}) - \alpha_0({\rm GS})$} &
				
				\multicolumn{1}{c}{$\delta\nu_{BBR}$[Hz]}&
				\multicolumn{1}{c}{$\nu$[Hz]}&
				\multicolumn{1}{c}{$\delta\nu_{BBR}$/$\nu$} \\
				\hline
				
				\multicolumn{7}{c}{\textbf{Sb~II}}\\
				
				2 $-$ 1  &21.641&22.387&-0.746&-0.6424[-2]&9.157$[13]$&$-7.02[-17]$ \\
				
				3 $-$ 1  &21.641&22.870&-1.229&-0.0106&$1.696[14]$&$-6.25[-17]$ \\
				
				\\
				
				\multicolumn{7}{c}{\textbf{Au~II}}\\
				
				2 $-$ 1 &12.400&34.737&-22.337&-0.1924&4.509[14]&$ -4.27[-16] $\\
				
				3 $-$ 1 &12.400&35.446&-23.046&-0.1985&5.289[14]&$ -3.75[-16] $\\
				4 $-$ 1 &12.400&34.519&-22.119&-0.1905&8.324[14]&$ -2.29[-16] $\\
				
				\\
				
				\multicolumn{7}{c}{\textbf{Hg~III}}\\
				2 $-$ 1 &7.376&19.444&-12.068&-0.1039&1.285[15]&$ -8.09[-17] $\\
				
				3 $-$ 1 &7.376&19.796&-12.420&-0.1070&1.380[15]&$ -7.75[-17] $\\

			\end{tabular}
		\end{ruledtabular}
	\end{table*}

In Table~\ref{t:CorPol}, we present our results of the core polarizability of the considered ions and compare them with experimental and other earlier calculations. As can be seen, the present results are in good agreement with those previous studies. 
We remind the reader that we perform the calculations for the Sb$ ^{+} $ ion in the $V^{N-4}$ approximation, i.e. the core of the ion is actually the  Sb$^{5+}$ ion.
	
In order to calculate the polarizabilities of states with open shells, we apply the approach developed in Ref.~\cite{dzuba2020calculation}. This approach is based on combining Eq.~(\ref{e:pol}) and the Dalgarno-Lewis method~\cite{dalgarno1955exact}, which reduces the summation of the complete set of states into the solution of a matrix equation. In our considered ions, all excited clock states and the ground state of the Sb$ ^{+} $ ion ($5s^25p^2$) are treated as open-shell systems.

	Our results for the scalar polarizabilities of the ground states and excited clock states for all considered ions are summarized in Table~\ref{t:pol}. Note that the values of GS and CS offered in this table indicate the total scalar polarizability (core + valence).
	The table illustrates that the polarizabilities of the clock states of the Sb$ ^{+} $ ion have similar values to the ground state. The reason for this is that both excited clock states are related to the same fine-structure  multiplet, and the energy intervals between them are considerably smaller than the excitation energies of the opposite-parity states. For the Au$ ^{+} $ and Hg$ ^{2+} $ ions, the difference in values of the polarizabilities is larger. It is due to the difference in electronic configurations between the ground and excited states; it is the $5d^{10}$ configuration in the ground state and the $5d^{9}6s$ configuration in the clock state.
That is similar to what occurred with the Yb$ ^{2+} $ ion in our previous study~\cite{allehabi2021using}.
	
The BBR shift of the optical clock frequencies can generally be described as the difference between the BBR shifts of the initial and final states in the transition, which is proportional to the difference in the scalar polarizabilities between the states. The BBR shift is given by (see, e.g., \cite{PhysRevA.74.020502})
\begin{equation}
	\delta \nu_{\mathrm{BBR}}=-8.611 \times 10^{-3}\left(\frac{T}{300 K}\right)^4 \Delta \alpha_0,
\end{equation}
where $ T  $ is a temperature in $ K $ (e.g., room-temperature $ T =
300 K $), $\Delta \alpha _0$= $\alpha_0({\rm CS}) - \alpha_0({\rm GS})$ is the difference between
the clock state and the ground-state polarizabilities presented
in atomic units. The BBR shifts for clock states considered in this work are shown in Table~\ref{t:pol}. As can be seen from the table, the BBR shifts for the Sb$ ^{+} $ and Hg$ ^{2+} $ ions are among the smallest, $\sim 10^{-17}$, while for the Au$^{+}$ ion, the relative BBR shifts are $\sim 10^{-16}$, which is similar to that for other atomic clocks.

One of the advantages of having at least two clock transitions in atoms or ions is that it is possible to combine the two frequencies of the clock transitions into one "synthetic" frequency which is not sensitive to the BBR shift~\cite{Yudin}. This trick can be used for any of the systems considered in the present work. 
The synthetic frequency is given by
\begin{equation}\label{e:syn}
	\omega_{ij} = \omega_i - \epsilon_{ij} \omega_j,
\end{equation}
where indexes $i$ and $j$ numerate clock transitions, $\epsilon_{ij} = \Delta \alpha_{0i}/\Delta \alpha_{0j}$. 
Using data from Table~\ref{t:pol} one can make one synthetic frequency for each of the ions, Sb~II and Hg~III, and three synthetic frequencies for the Au~II ion. See section \ref{s:alpha} for further discussion.

\subsection{ Electric quadrupole moments}
\label{s:Q}	
Clock states of the considered ions have relatively large values of the total angular momentum ($J=2$ or $3$).
This means that they are affected by the interaction of the quadrupole moments of the states with the gradient of the external electric field leading to the shift of frequency. Therefore, it is important to know the values of the quadrupole moments.
	
The hamiltonian of the interaction can be written as
\begin{equation}
	H_{\mathcal{Q}}=\sum_{q=-1}^{1}(-1)^{q} \nabla \mathcal{E}_{q}^{(2)} \hat \Theta_{-q}.
\end{equation}
Here, the tensor $ \nabla \mathcal{E}_{q}^{(2)}$ represents the external electric field gradient at the atom's position, and $\hat \Theta_{q}$ describes the electric-quadrupole operator for the atom. It is the same as for the $E2$ transitions, $\hat \Theta_q = r^2C^{(2)}_q$, where $C^{(2)}_q$ is the normalized spherical function and $q$ indicates the operator component.


	\begin{table}[!]
		\caption{\label{t:Q}
			Quadrupole moment ($\Theta$, a.u.) of the considered optical clock states.} 
		
		\begin{ruledtabular}
			\begin{tabular}{crccccc}
				
				\multicolumn{1}{c}{No.}& 
				\multicolumn{1}{c}{Conf.}&
				\multicolumn{1}{c}{Term}&
				\multicolumn{1}{c}{$J$}&
				
				\multicolumn{1}{c}{$E$ (cm$^{-1}$)}&
				
				\multicolumn{1}{c}{ ME (a.u.) }&
				\multicolumn{1}{c}{ $\Theta$ }\\&&&&&

				\multicolumn{1}{c}{ $\left\langle J\|\Theta_{0}\| J\right\rangle$}\\
				\hline
				\multicolumn{7}{c}{\textbf{Sb~II}}\\
				2 & $5s^25p^2$& $^3${P}& {1} &3054.6&  -0.176& -0.032\\
				3& $5s^25p^2$& $^3${P}& {2} &5658.2& 0.092& 0.022\\

				\\
				
				\multicolumn{7}{c}{\textbf{Au~II}}\\
				
				2& $5d^{9}6s$& $^3${D}& {3}&15039.572&3.497&0.853\\
				
				3& $5d^{9}6s$& $^3${D}& {2}&17640.616&2.452&0.586\\
				
				4& $5d^{9}6s$& $^3${D}& {1}&27765.758&1.488&0.272\\
				\\
				
				\multicolumn{7}{c}{\textbf{	Hg III}}\\
				
				2& $ 5d^{9}6s $  & (5/2,1/2)  & 3 &42850.3&   2.829& 0.690\\ 
				3& $ 5d^{9}6s $  & (5/2,1/2)  & 2&46029.5&2.004& 0.479 \\ 
				
			\end{tabular}
			
		\end{ruledtabular}
		
	\end{table}
	
	\begin{equation}
		H_{\mathcal{Q}}=\sum_{q=1}^{-1}(-1)^{q} \nabla \mathcal{E}_{q}^{(2)} \hat \Theta_{-q}.
	\end{equation}
The electric quadrupole moment $\Theta$ can be calculated as the expectation value of the $\hat\Theta_{0}$ operator
\begin{equation}
	\begin{aligned}
		\Theta &=\left\langle n J J\left|\hat \Theta_{0}\right| n J J\right\rangle\\
		&=\left\langle n J\|\hat\Theta\| n J\right\rangle \sqrt{\frac{J\left(2 J-1\right)}{\left(2 J+3\right)\left(2 J+1\right)\left(J+1\right)}},
	\end{aligned}
\end{equation}
where $\left\langle n J\|\hat\Theta\| n J\right\rangle$ denotes the reduced matrix element of the electric quadrupole operator.
	
	
Table~\ref{t:Q} shows the reduced ME of the electric quadrupole operator and their quadrupole moments for the states considered. 
It should be noted that since the total angular momentum $J$ of the ground states of all atomic systems is 0, no electric quadrupole operator ($\Theta = 0$) exists for them. 

The quadrupole shift can be suppressed by averaging over transitions between different hyperfine or Zeeman components~\cite{dzuba2018testing,itano2000external}. For bosonic isotopes of Hg$^{2+}$ the shift can be avoided for the clock state with $J=3$ by using Zeeman components with $J_z= \pm 2$. This is because $\Delta E \sim 3J_z^2-J(J+1)$.

\subsection{Sensitivity of the clock transitions to the variation of the fine-structure constant}

\label{s:alpha}
	
	There has been considerable interest in the use of optical atomic clock transitions for the search for  the time variation of the fine-structure constant ($\alpha$)~\cite{dzuba1999space,dzuba2009atomic,dzuba1999calculations,flambaum2009search}. The frequencies of these transitions depend differently on $\alpha$. Therefore, if $\alpha$ changes the frequency ratio also changes. These can be revealed by repeated measurements over a long period of time.
	
It is convenient to present the dependence of atomic frequency on $\alpha$ in the vicinity of its physical value in a form	
\begin{equation}\label{e:q}
	\omega = \omega_0 + q\left[\left(\frac{\alpha}{\alpha_0}\right)^2-1\right],
\end{equation}
where $\alpha_0$ and $\omega_0$ are the laboratory values of the fine structure constant and the transition frequency, respectively, and $q$ is the sensitivity coefficient which is determined from atomic calculations~\cite{dzuba1999space,dzuba1999calculations,flambaum2009search}.  
Calculations are done for at least two different values of $\alpha$ and then the numerical derivative is taken
\begin{equation}
	q=\frac{\omega(x)-\omega(-x)}{2x},
\end{equation}
where $x=(\alpha/\alpha_0)^2-1$ [see Eq.~(\ref{e:q})]. It is imperative to point out that the $x$ value must be as small as possible in order to achieve linear behaviour. At the same time, it must be large enough to suppress numerical noise. In most cases, $x=0.01$ is adequate for obtaining accurate results.
	
The change in frequencies ratio can be linked to the change of $\alpha$ by the formula
\begin{equation}
	\delta\left(\frac{\omega_1}{\omega_2}\right)=
	\frac{\delta\omega_1}{\omega_1} - \frac{\delta\omega_2}{\omega_2} = \left(K_1 - K_2 \right)\frac{\delta\alpha}{\alpha}.
\end{equation}
Here, $K=2q/\omega$, and it is often referred to as the enhancement factor. A summary of the calculated $q$ and $K$ for all considered clock transitions is given in Table~\ref{t:q}. It can be seen that the enhancement coefficient $K$ is large ($K>2$) for all studied clock transitions. 
Note that both clock transitions in Sb$^+$ are transitions within the same fine structure (FS) multiplet. Assuming $\Delta E_{\rm FS} \sim (Z\alpha)^2$, which is often true to high precision for light elements, leads to $K=2$. The fact that $K>2$ means that higher in  $(Z\alpha)$ terms give  a significant contribution to the FS.

Clock transitions in Au$^{+}$ and Hg$^{2+}$ have a different nature. They are the $s-d$ transitions in single-electron approximation. This ensures high sensitivity to the $\alpha$ variation~\cite{dzuba1999space}. Note that the values of the sensitivity coefficients $q$ are more important than the values of the enhancement factors $K$~\cite{dzuba2018testing}. The values of $q$ for clock transitions in Au$^{+}$ and Hg$^{2+}$ are among the highest considered so far. This is especially true for Hg$^{2+}$. Only clock transitions in Yb$^+$ and Hg$+$ have larger or similar values of $q$~\cite{dzuba2018testing}.
Both these ions were already used in the search for time variation of $\alpha$~\cite{PhysRevLett.113.210801,Science.319}.

		\begin{table}[!]
		\caption{\label{t:q}
			 The sensitivity of the clock transitions to variation of the fine-structure constant ($q, K$). The theoretical value of the transition frequency $\omega$ is used to calculate $K$ ($K=2q/\omega$).} 
		
		\begin{ruledtabular}
			\begin{tabular}{crccccc}
				
				\multicolumn{1}{c}{No.}& 
				\multicolumn{1}{c}{Conf.}&
				\multicolumn{1}{c}{Term}&
				\multicolumn{1}{c}{$J$}&
				
				\multicolumn{1}{c}{$\omega$ (cm$^{-1}$)}&
				
				\multicolumn{1}{c}{ $q$ (cm$^{-1}$) }&
				\multicolumn{1}{c}{$K$}\\

				\hline
				\multicolumn{7}{c}{\textbf{Sb~II}}\\
				2 & $5s^25p^2$& $^3${P}& {1} &3146 & 4382 & 2.79\\
				3& $5s^25p^2$& $^3${P}& {2} & 5504 & 6060 & 2.20\\
				
				\\
				
				\multicolumn{7}{c}{\textbf{Au II}}\\

				2& $5d^{9}6s$& $^3${D}& {3}&15095&-41950&-5.56\\
				
				3& $5d^{9}6s$& $^3${D}& {2}&17832&-40109&-4.50\\
				
				4& $5d^{9}6s$& $^3${D}& {1}&27225&-30793&-2.26\\
				
				\\
				
				\multicolumn{7}{c}{\textbf{	Hg III}}\\
                                
				2& $ 5d^{9}6s $  & (5/2,1/2)  & 3 &43360&-54845&-2.53 \\ 
				3& $ 5d^{9}6s $  & (5/2,1/2)  & 2&46616&-52807&-2.27   \\ 
				
			\end{tabular}
			
		\end{ruledtabular}
		
	\end{table}

It was stated in section \ref{s:BBR} that combining two clock frequencies with appropriate coefficient one can make a synthetic frequency, which is not sensitive to the BBR shift (see also Ref.~\cite{Yudin}). Such frequencies are still sensitive to the variation of $\alpha$, 
\begin{equation}\label{e:dotnu}
	\frac{\dot \omega_{ij}}{\omega_{ij}} = \frac{K_i\omega_i - \epsilon_{ij}K_j\omega_j}{\omega_i-\epsilon_{ij}\omega_j}\frac{\dot \alpha}{\alpha} \equiv K_{ij}\frac{\dot \alpha}{\alpha}.
\end{equation}
The values of synthetic frequencies for Sb~II, Au~II and Hg~III and corresponding enhancement factors $K_{ij}$ are presented in Table~\ref{t:Kij}.
Note that the values of $K_{ij}$ are relatively large and have an opposite sign to the enhancement factors for frequencies of direct clock transitions.
It might be possible to compare a synthetic frequency to a direct frequency for higher sensitivity to the variation of $\alpha$.

\begin{table}[!]
	\caption{\label{t:Kij}
		Synthetic frequencies of Sb~II, Au~II and Hg~III clock transitions and their sensitivity to variation of the fine-structure constant.
			Indexes $i$ and $j$ correspond to the clock transitions from state number $i$ or $j$ (see Table~\ref{t:Energy}) to the ground state.} 
		\begin{ruledtabular}
			\begin{tabular}{l cccrr}
				\multicolumn{1}{c}{Ion}& 
				\multicolumn{1}{c}{$i$}& 
				\multicolumn{1}{c}{$j$}& 
				\multicolumn{1}{c}{$\epsilon_{ij}$}& 
				\multicolumn{1}{c}{$\omega_{ij}$ [cm$^{-1}$]}& 
				\multicolumn{1}{c}{$K_{ij}$}\\
				\hline 
Sb~II   & 2 &  1 &    1.647 &  625.890  &  -2.544 \\
\\
Au~II   & 2  & 1 &     1.032 & 2123.672  &   3.245 \\
            & 3  & 1 &    0.990  & 12872.976 &    1.558 \\
            &  3  & 2 &   0.960 & 10834.726  &   1.240 \\
            \\
Hg~III &   2  & 1 &  1.029  & 1929.340  &   3.673 \\
			\end{tabular}
		\end{ruledtabular}
	\end{table}

\subsection{Search for new interactions using non-linearities of King plot.}	

It is very well known that the study of isotope shift (IS) in atomic transitions can reveal  important information on nuclei, e.g., the change of nuclear radius between two isotopes. For many purposes, it is convenient to combine the data into the so-called King plot. King plot represents the IS shift of one transition (modified by reduced mass of the pair of isotopes) as a function of the modified IS of another transition. Under normal conditions, the King plot is a straight line. Deviations from a straight line (the so call non-linearities of King plot) may reveal further information about nuclei. It was suggested in Refs.~\cite{KPN1,KPN2} that new electron-neutron interaction may lead to non-linearities of the King plot due to a different number  of  neutrons in different isotopes. Therefore, the study of the non-linearities may put  a limit on the strength of such interaction. It was further noted in Ref.~\cite{deform} that the non-linearities may originate from nuclear deformation, opening another way of studying nuclear shape.

The minimum requirements to study non-linearities of King plot include having two atomic transitions, four isotopes and high accuracy of the measurements.
The  latter requirement means that using clock transitions is the best choice. Among the systems considered in the present work only Hg$^{2+}$ satisfies all the requirements. Hg has seven stable isotopes, and five of them have zero nuclear spin. Having zero nuclear spin is preferable for IS measurements to avoid complications caused by the hyperfine structure. On the other hand, the 2 to 1 clock transition in Hg$^{2+}$ becomes very weak in the absence of the hyperfine structure. This can be probably resolved by applying  a strong magnetic field. It would play a role similar to magnetic hfs, the transition rate is given by Eq.~(\ref{e:AE2}) in which the operator of magnetic hfs is replaced by M1 operator times the strength of the magnetic field.

Applying a strong magnetic field would lead to Zeeman shift of the transition frequencies. However, first-order Zeeman shift can be avoided by working with zero projection of the total momentum $J$, or by averaging the frequencies for $\pm J_z$. The latter option with $J_z = \pm 2$ is good for simultaneous cancelation of Zeeman and quadrupole shifts (see section \ref{s:Q}). Second-order Zeeman shift is usually small in the absence of hyperfine structure.
	
	\section{Summary}

We concluded that the metastable states of Sb$ ^{+} $, Au$ ^{+} $, and Hg$ ^{2+} $ ions could make good optical clocks with the advantage of being sensitive to fine-structure constant variations. 
Several atomic properties relevant to the development of optical clocks have been calculated using relativistic many-body methods. It was found that the relative BBR shifts for the transitions $\sim 10^{-16}$ and can be further suppressed by constructing synthetic frequencies. 
The sensitivity of the clock transitions to the variation of the fine structure constant is found to be one of the highest among operating or proposed clock transitions. 
All considered systems have two or three clock transitions which bring extra benefits for suppressing BBR shift by combining the frequencies into the so-called synthetic frequencies in which the BBR shift cancels out; or for monitoring the time variation of the fine structure constant by measuring one frequency against the other many times over  a long period of time. In addition, clock transitions in the Hg$^{2+}$ ion can be used to put  a limit on new electron-neutron interaction or to study nuclear shape via measuring non-linearities of King plot.

	\begin{acknowledgments}
	This work was supported by the Australian Research Council Grants No. DP190100974 and DP200100150. S.O.A. gratefully acknowledges the Islamic University of Madinah (Ministry of Education, Kingdom of Saudi Arabia) for funding his scholarship.  This research includes computations using the computational cluster Katana supported by Research Technology Services at UNSW Sydney~\cite{Katana}.
	\end{acknowledgments}

	\bibliography{references}

\end{document}